\RequirePackage{fix-cm}

\documentclass[smallcondensed
]{svjour3}     % onecolumn (ditto)
\smartqed  % flush right qed marks, e.g. at end of proof
\usepackage{graphicx}
\usepackage{epstopdf}
\usepackage{geometry}
\usepackage{comment}
\usepackage{xspace}
\usepackage{amsmath}
\usepackage{amssymb}

\usepackage{epsfig}
\usepackage{subfigure}
\usepackage{algorithmic}
\usepackage{graphics}
\usepackage{alltt}
\usepackage{float}
\usepackage{array}
\usepackage{footmisc}
\usepackage{comment}
\usepackage{algorithm}
\usepackage{url}
\usepackage[misc]{ifsym}

\usepackage[justification=centering]{caption}
\begin{document}

\title{Efficient Multimedia Similarity Measurement Using Similar Elements%\thanks{Grants or other notes
}


\author{Chengyuan Zhang  $^\dagger$\and
        Yunwu Lin $^\dagger$\and
        Lei Zhu $^\dagger$\and
        Zuping Zhang $^\dagger$\and
        XinPan Yuan $^\ddagger$\and
        Fang Huang $^\dagger$\and
        }


\institute{Chengyuan Zhang \at
            \email{cyzhang@csu.edu.cn}
            \and
            Yunwu Lin \at
              \email{lywcsu@csu.edu.cn}
           \and
           \Letter Lei Zhu \at
              \email{leizhu@csu.edu.cn}
           \and
           Zuping Zhang \at
              \email{zpzhang@csu.edu.cn}
           \and
           XinPan Yuan \at
              \email{xpyuan@hut.edu.cn}
           \and
           Fang Huang \at
              \email{hfang@csu.edu.cn}
           \and
           $^\dagger$ School of Information Science and Engineering, Central South University, PR China\\
           $^{\ddagger}$ School of Computer, Hunan University of Technology, China\\
}
\begin{comment}
\institute{Chengyuan Zhang \at
            \email{cyzhang@csu.edu.cn}
            \and
            Yunwu Lin \at
              \email{lywcsu@csu.edu.cn}
           \and
           \Letter Lei Zhu \at
              \email{leizhu@csu.edu.cn}
           \and
           Zuping Zhang \at
              \email{zpzhang@csu.edu.cn}
           \and
           Yan Tang \at
              \email{tangyan@csu.edu.cn}
           \and
           Fang Huang \at
              \email{hfang@csu.edu.cn}
           \\
           \\
           $^\dagger$ School of Information Science and Engineering, Central South University, PR China\\
}
\end{comment}

\date{Received: date / Accepted: date}

\maketitle

\begin{abstract}
online social networking techniques and large-scale multimedia systems are developing rapidly, which not only has brought great convenience to our daily life, but generated, collected, and stored large-scale multimedia data. This trend has put forward higher requirements and greater challenges on massive multimedia data retrieval. In this paper, we investigate the problem of image similarity measurement which is used to lots of applications. At first we propose the definition of similarity measurement of images and the related notions. Based on it we present a novel basic method of similarity measurement named SMIN. To improve the performance of calculation, we propose a novel indexing structure called SMI Temp Index (SMII for short). Besides, we  establish an index of potential similar visual words off-line to solve to problem that the index cannot be reused. Experimental evaluations on two real image datasets demonstrate that our solution outperforms state-of-the-art method.


\end{abstract}

\section{Introduction}
\label{intro}

In the recent years, online social networking techniques and large-scale multimedia systems~\cite{DBLP:conf/cikm/WangLZ13,DBLP:conf/mm/WangLWZZ14,DBLP:conf/mm/WangLWZ15,DBLP:journals/tip/WangLWZZH15} are developing rapidly, which not only has brought great convenience to our daily life, but generated, collected, and stored large-scale multimedia data~\cite{DBLP:journals/tip/WangLWZ17}, such as text, image~\cite{DBLP:conf/sigir/WangLWZZ15}, audio, video~\cite{LINYANGARXIV} and 3D data. For example, in China, Weibo (https://weibo.com/) is the largest online social networking service, which have 376 million active users and more than 100 million micro-blogs containing short text, image, or short video are posted. The most famous social networking platform all over the world, Facebook (https://facebook.com/), reports 350 million images uploaded everyday in the end of November 2013. More than 400 million tweets with texts and images have been generated by 140 million users on Twitter (http://www.twitter.com/) which is another popular social networking web site in the world. In September 2017, the largest online social networking platform. Another type of common application in the Internet is multimedia data sharing services. Flickr(https://www.flickr.com/) is one of the most famous photos sharing web site around the world. More than 3.5 million new images uploaded to this platform everyday in March 2013. More than 14 million articles are clicked every day on Pinterest, which is an attractive image social networking web site. More than 2 billion totally videos stored in YouTube, the most famous video sharing platform by the end of 2013, and every minute there are 100 hours of videos which are uploaded to this service (https://www.youtube.com/). The total watch time exceeded 42 billion minutes on IQIYI (http://www.iqiyi.com/), the most famous online video sharing service in China and number of independent users monthly is more than 230 million monthly. For audio sharing services, the total amount of audio in Himalaya (https://www.ximalaya.com/) had exceeded 15 million as of December 2015. Other web services like Wikipedia (https://en.wikipedia.org/), the largest and most popular free encyclopedia on the Internet, contains more than 40 million articles with pictures in 301 different languages. Other mobile applications such as WeChat, Instagram, etc, provide great convenience for us to share multimedia data. Thanks to these current rich multimedia services and applications, multimedia techniques~\cite{DBLP:conf/ijcai/WangZWLFP16,DBLP:journals/cviu/WuWGHL18} is changing every aspect of our lives. On the other hand, the emergence of massive multimedia data~\cite{DBLP:journals/corr/abs-1708-02288} and applications puts forward greater challenges for techniques of information retrieval.

\begin{figure*}
\newskip\subfigtoppskip \subfigtopskip = -0.1cm
%\newskip\subfigcapskip \subfigcapskip = -0.2cm
\begin{minipage}[b]{0.99\linewidth}
\begin{center}
   %  \centering
     \includegraphics[width=1\linewidth]{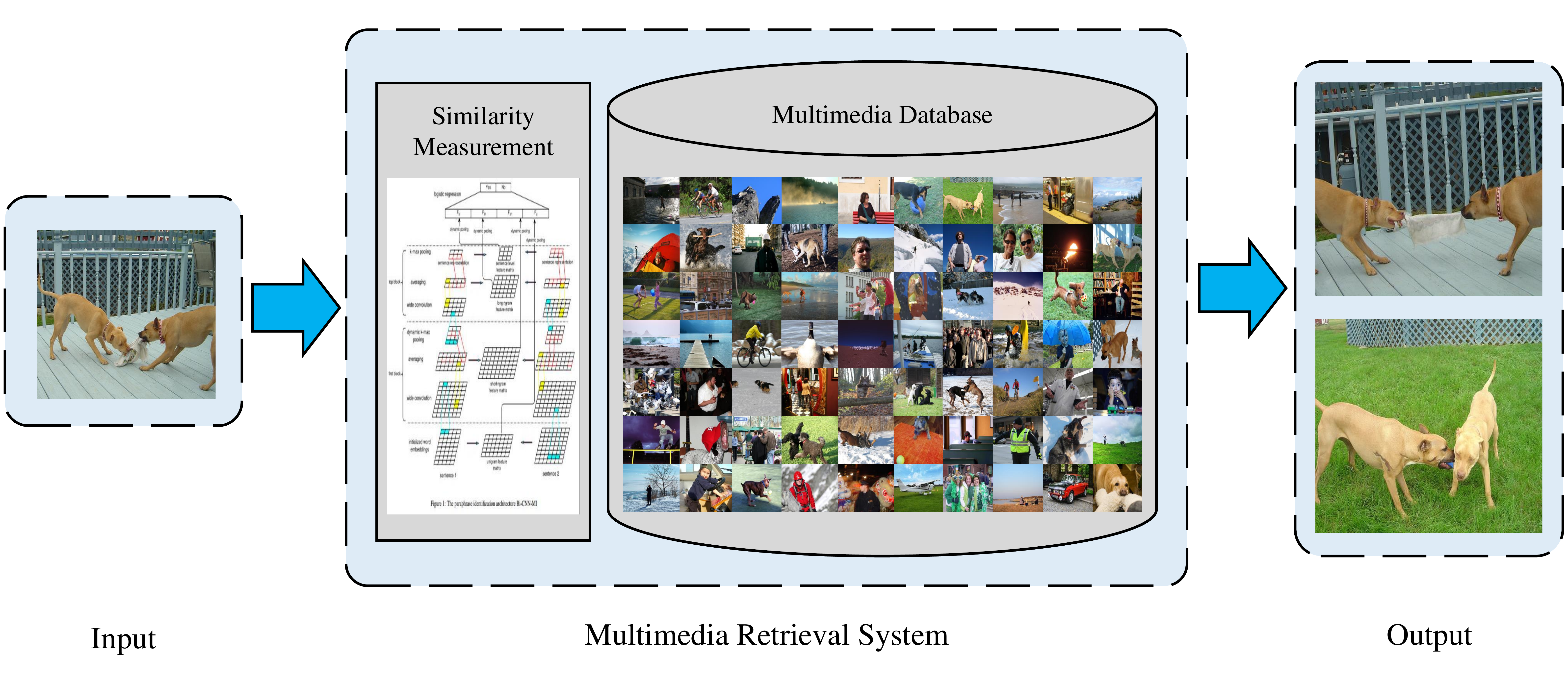}
   %\captionsetup{justification=centering,margin=2cm}
   \captionsetup{justification=centering}
       \vspace{-0.2cm}
\caption{An example of multimedia retrieval via similarity measurement}
\label{fig:fig1}
\end{center}
\end{minipage}
\label{fig:k}
\end{figure*}

\begin{figure*}
\newskip\subfigtoppskip \subfigtopskip = -0.1cm
%\newskip\subfigcapskip \subfigcapskip = -0.2cm
\begin{minipage}[b]{0.99\linewidth}
\begin{center}
   %  \centering
     \includegraphics[width=1\linewidth]{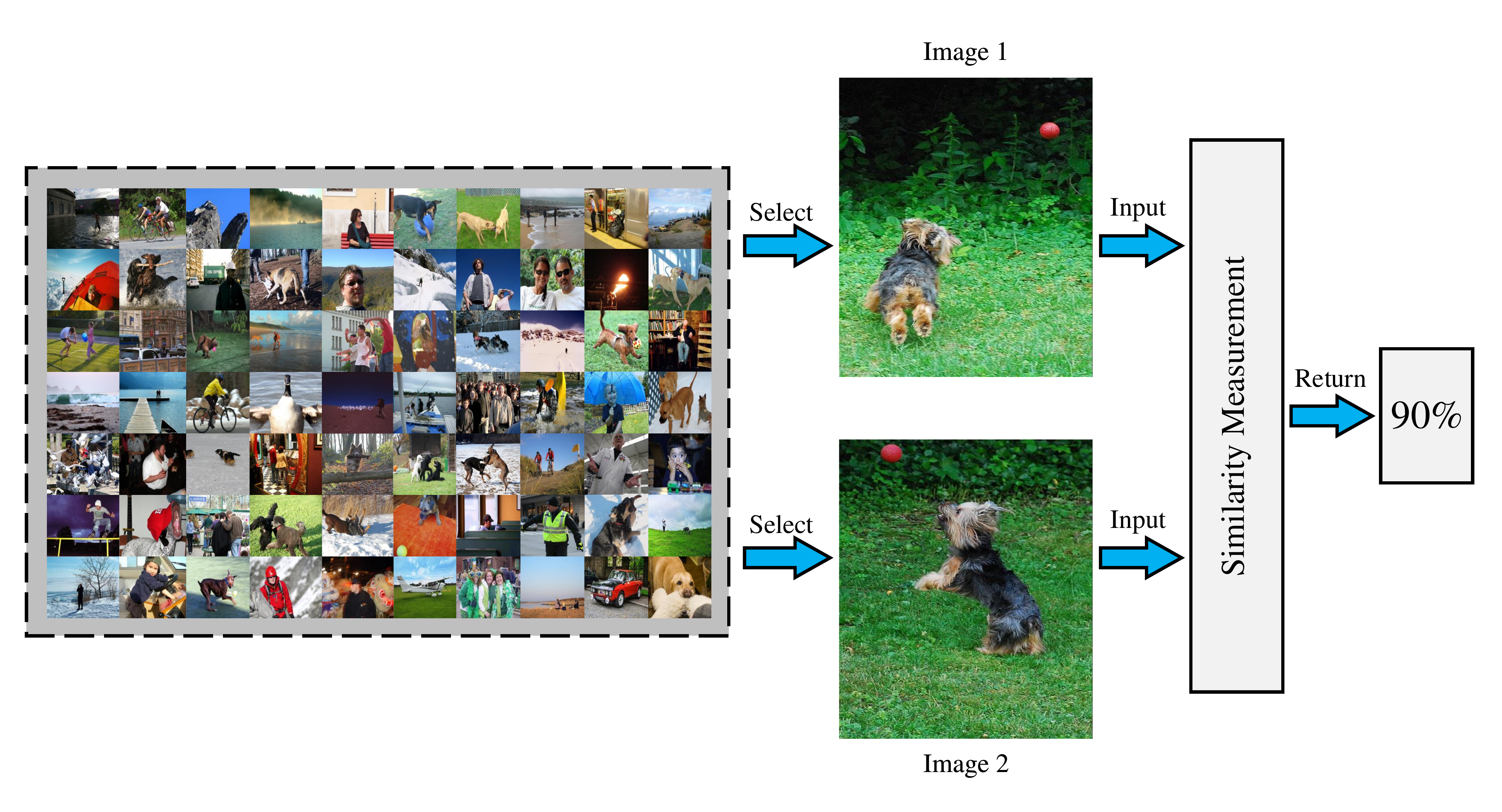}
   %\captionsetup{justification=centering,margin=2cm}
   \captionsetup{justification=centering}
       \vspace{-0.2cm}
\caption{An example of multimedia retrieval via similarity measurement}
\label{fig:fig2}
\end{center}
\end{minipage}
\label{fig:k}
\end{figure*}

\noindent\textbf{Motivation}. Textual similarity measurement is a classical issue in the community of information retrieval and data mining, and lots of approaches have been proposed. Guo et al~\cite{Guo2016Sentence} proposed to use vectors as basic elements, and the edit distance and Jaccard coefficient are used to calculate the sentence similarity. Li et al.~\cite{LiFeng2017} proposed the use of word vectors to represent the meaning of words, and considers the influence of multiple factors such as word meaning, word order and sentence length on the calculation of sentence similarity. Unlike the studies of textual similarity measurement, in this paper we investigate the problem of image similarity measurement, which is a widely applied technique in lots of application scenarios such as image retrieval~\cite{DBLP:conf/pakdd/WangLZW14,DBLP:journals/pr/WuWGL18,DBLP:journals/tnn/WangZWLZ17}, image similarity calculation and matching~\cite{NNLS2018,TC2018}. There are two examples shown in Figure~\ref{fig:fig1} and Figure~\ref{fig:fig2} which can describe this problem in a more clear way.

\begin{example}
\label{ex:example1}
In Figure~\ref{fig:fig1}, An user have a photo and she want to find out others pictures which are highly similar to it in the Internet. She can submit a image query containing this photo into the multimedia retrieval system. The system measures the visual content similarity between this photo and the images in the database and after that a set of similar images is returned.
\end{example}

\begin{example}
\label{ex:example2}
Figure.~\ref{fig:fig2} demonstrates another example of application of image similarity measurement. An user want to measure similarity betweeen two pictures in a dataset quantitatively. She selects two pictures from the image dataset and input them into the similarity measurement system. According to image similarity measurement algorithm, the system will calculate the value of similarity between these images (e.g., 90\%).
\end{example}

In order to improve the efficiency and accuracy of image similarity measurement, we present the definition of similarity measurement of images and the relevant notions. We introduce the measurement of similar visual words named SMI Naive (SMIN for short) which is the basic method for similarity measurement, and then propose the SMIN algorithm. After that, to optimize this method, we design a novel indexing structure named SMI Temp Index to reduce the time complexity of calculation. In addition, another technique named index of potential similar visual words is proposed to solve the problem that the index cannot be reused. We could search for the index to perform the measurement of similar visual words without having to repeatedly create a temporary index.

\noindent\textbf{Contributions}. Our main contributions can be summarized as follows:
\begin{itemize}
\item  Firstly we introduce the definition of similarity measurement of images and the related conceptions. The image similarity calculation function are designed.
\item We introduce the basic method of image similarity measurement, called SMI Naive (SMIN for short). In order to improve the performance of similarity measurement, based on it we design two indexing techniques named SMI Temp Index (SMII for short) and Index of Potential Similar Visual Words (PSMI for short).
\item We have conducted extensive experiments on two real image datasets. Experimental results demonstrate that our solution outperforms the state-of-the-art method.
\end{itemize}

\noindent\textbf{Roadmap.} In the remainder of this paper, Section~\ref{relwork} presents the related works about image similarity measurement and image retrieval. In Section~\ref{preliminaries}, the definition of image similarity measurement and related conceptions are proposed. We present the basic similarity measurement method named SMIN and two improved indexing techniques and algorithms in Section~\ref{optimization}. Our experimental results are presented in Section~\ref{perform}. Finally, we conclude the paper in Section~\ref{con}.

\section{Related Work}
\label{relwork}

In this section, we present the related works of image similarity measurement and image retrieval, which are relevant to this study.

\noindent\textbf{Image Similarity Measurement.} In recent years, image similarity measurement has become a hot issue in the community of multimedia system~\cite{DBLP:journals/ivc/WuW17} and information retrieval since the massive image data can be accessed in the Internet. On the other hand, like textual similarity measurement, image similarity measurement is an important technique which can be applied in lots of applications, such as image retrieval, image matching, image recognition and classification, computer vision, etc. Many researchers work for this issue and numerous approaches have been proposed. For example, Coltuc et al.~\cite{DBLP:journals/entropy/ColtucDC18} studied the usefulness of the normalized compression distance (NCD for short) for image similarity detection. In their work, they considered correlation between NCD based feature vectors extracted for each image. Albanesi et al.~\cite{DBLP:journals/jmiv/AlbanesiABM18} proposed a novel class of image similarity metrics based on a wavelet decomposition. They investigated the theoretical relationship between the novel class of metrics and the well-known structural similarity index. Abe et al.~\cite{DBLP:conf/iccae/AbeMH18} studied similarity retrieval of trademark images represented by vector graphics. To improve the performance of the system, they introduced centroid distance into the feature extraction. Cicconet et al.~\cite{DBLP:journals/corr/abs-1802-06515} studied the problem of detecting duplication of scientific images. They introduced a data-driven solution based on a 3-branch Siamese Convolutional Neural Network which can serve to narrow down the pool of images. For multi-label image retrieval, Zhang et al.~\cite{DBLP:journals/corr/abs-1803-02987} proposed a novel deep hashing method named ISDH in which an instance-similarity deﬁnition was applied to quantify the pairwise similarity for images holding multiple class labels. Kato et al.~\cite{DBLP:journals/ipsjtcva/KatoSP17} proposed a novel solutions for the problem of selecting image pairs that are more likely to match in Structure from Motion. They used Jaccard Similarity and bag-of-visual-words in addition to tf-idf to measure the similarity between images. Wang et al~\cite{Wang2016Semantic} designed a regularized distance metric framework which is named semantic discriminative metric learning (SDML for short). This framework combines geometric mean with normalized divergences and separates images from different classes simultaneously. Guha et al.~\cite{Guha2013Image} proposed a new approach called Sparse SNR (SSNR for short) to measuring the similarity between two images using sparse reconstruction. Their measurement does not need to use any prior knowledge about the data type or the application. KHAN et al.~\cite{DBLP:journals/ieicet/KhanAT12} proposed two halftoning methods to improve efficiency in generating structurally similar halftone images using Structure Similarity Index Measurement. Their Method I can improves efficiency as well as image quality and Method II can reaches a better image quality with fewer evaluations than pixel-swapping algorithm used in Method I.

Near-duplicate image detection is a another problem related to image similarity measurement. To solve the problem of near-duplicate image retrieval, Wang et al.~\cite{DBLP:journals/ijes/WangZ18} developed a novel spatial descriptor embedding method which encodes the relationship of the SIFT dominant orientation and the exact spatial position between local features and their context. Gadeski et al.~\cite{DBLP:journals/mta/ZhaoLPF17} proposed an effective algorithm based on MapReduce framework to identify the near duplicates of images from large-scale image sets. Nian et al.~\cite{DBLP:journals/mta/NianLWGL16} investigated this type of problem and presented an effective and efficient local-based representation method named Local-based Binary Representation to encode an image as a binary vector. Zlabinger et al.~\cite{DBLP:conf/sac/ZlabingerH17} developed a semi-automatic duplicate detection approach in which single-image-duplicates are detected between sub-images based on a connected component approach and duplicates between images are detected by using min-hashing method. Hsieh et al.~\cite{DBLP:journals/mta/HsiehCC15} designed a novel framework that adopts multiple hash tables in a novel way for quick image matching and efficient duplicate image detection. Based on a hierarchical model, Li et al.~\cite{DBLP:journals/mta/LiQLZWT15} introduced an automatic NDIG mining approach by utilizing adaptive global feature clustering and local feature refinement to solve the problem of near duplicate image groups mining. Liu et al.~\cite{DBLP:journals/tip/LiuLS15} presented a variable-length signature to address the problem of near-duplicate image matching. They used the earth mover's distance to handle variable-length signatures. Yao et al.~\cite{DBLP:journals/spl/YaoYZ15} developed a novel contextual descriptor which measures the contextual similarity of visual words to immediately discard the mismatches and reduce the count of candidate images. For large scale near-duplicate image retrieval Fedorov et al.~\cite{DBLP:journals/corr/FedorovK16} introduced a feature representation combining of three local descriptors, which is reproducible and highly discriminative. To improve the efficiency of near-duplicate image retrieval, Yıldız et al.~\cite{DBLP:conf/ssiai/YildizD16} proposed a novel interest point selection method in which the distinctive subset is created with a ranking according to a density map.

\noindent\textbf{Image Retrieval.} Content-based image retrieval (CBIR for short)~\cite{DBLP:journals/pami/JingB08,DBLP:journals/tomccap/LewSDJ06,DBLP:conf/mm/WuWS13} is to retrieve images by analyzing visual contents, and therefore image representation~\cite{DBLP:conf/mm/WanWHWZZL14,DBLP:journals/pr/WuWGL18} plays an important role in this task. In recent years, the task of CBIR has attracted more and more attentions in the multimedia~\cite{TC2018,DBLP:journals/pr/WuWLG18} and computer vision community~\cite{DBLP:journals/tnn/WangZWLZ17,NNLS2018}. Many techniques have been proposed to support efficient multimedia query and image recognition. Scale Invariant Feature Transform (SIFT for short)~\cite{DBLP:conf/iccv/Lowe99,DBLP:journals/ijcv/Lowe04} is a classical method to extract visual features, which transforms an image into a large collection of local feature vectors. SIFT includes four main step: (1)scale-space extrema detection; (2)keypoint localization; (3)orientation assignment; (4)Kkeypoint descriptor. It is widely applied in lots of researches and applications. For example, Ke et al.~\cite{DBLP:conf/cvpr/KeS04} proposed a novel image descriptor named PCA-SIFT which combines SIFT techniques and principal components analysis (PCA for short) method. Mortensen et al.~\cite{DBLP:conf/cvpr/MortensenDS05} proposed a feature descriptor that augments SIFT with a global context vector. This approach adds curvilinear shape information from a much larger neighborhood to reduce mismatches. Liu et al.~\cite{DBLP:journals/inffus/LiuLW15} proposes a novel image fusion method for multi-focus images with dense SIFT. This dense SIFT descriptor can not only be employed as the activity level measurement, but also be used to match the mis-registered pixels between multiple source images to improve the quality of the fused image. Su et al.~\cite{Su2017MBR} designed a horizontal or vertical mirror reflection invariant binary descriptor named MBR-SIFT to solve the problem of image matching. Nam et al.~\cite{DBLP:journals/mta/NamKMHCL18} introduced a SIFT features based blind watermarking algorithm to address the issue of copyright protection for DIBR 3D images. Charfi et al.~\cite{DBLP:journals/mta/CharfiTAS17} developed a bimodal hand identification system based on SIFT descriptors which are extracted from hand shape and palmprint modalities.

Bag-of-visual-words~\cite{DBLP:conf/iccv/SivicZ03,DBLP:journals/tnn/WangZWLZ17,DBLP:journals/corr/abs-1804-11013}(BoVW for short) model is another popular technique for CBIR and image recognition, which was first used in textual classification. This model is a technique to transform images into sparse hierarchical vectors by using visual words, so that a large number of images can be manipulated. Santos et al.~\cite{DBLP:journals/mta/SantosMST17} presented the first ever method based on the signature-based bag of visual words (S-BoVW for short) paradigm that considers information of texture to generate textual signatures of image blocks for representing images. Karakasis et al.~\cite{DBLP:journals/prl/KarakasisAGC15} presents an image retrieval framework that uses affine image moment invariants as descriptors of local image areas by BoVW representation. Wang et al.~\cite{DBLP:conf/mmm/WangWLZ13} presented an improved practical spatial weighting for BoV (PSW-BoV for short) to alleviate this effect while keep the efficiency.

\section{Preliminaries}
\label{preliminaries}
In this section, we propose the definition of region of visual interests (RoVI for short) at the first time, then present the notion of region of visual interests query (RoVIQ for short) and the similarity measurement. Besides, we review the techniques of image retrieval which is the base of our work. Table~\ref{tab:notation} summarizes the notations frequently used throughout this paper to facilitate the discussion.

\begin{table}%\label{tab:symbols}
	\centering
    \small
	\begin{tabular}{|p{0.27\columnwidth}| p{0.70\columnwidth} |}
		\hline
		\textbf{Notation} & \textbf{Definition} \\ \hline\hline
		~$\mathcal{D}_I$                                 & A given database of images              \\ \hline
        ~$\mathcal{I}_i$                                 & The $i$-th image                 \\ \hline
        ~$\mathcal{W}_i$                                 & A visual words set                 \\ \hline
        ~$\mathcal{|W|}$                                 & The number of visual words in $\mathcal{W}$                \\ \hline
        ~$w^i_1$                                         & The 1-th visual word in the visual words set $\mathcal{W}_i$\\ \hline
        ~$\lambda_k$                                     & The similarity of two visual words  \\ \hline
	  	~$\mathcal{P}_k = (w^i_k,w^j_k)$                 & The similar visual word pair   \\ \hline
        ~$\bigotimes$                                    & The operator to generates the set of SVWPs      \\ \hline
        ~$\hat{\lambda}$                                 & The similarity threshold of predefined        \\ \hline
        ~$\Xi_i$                                         & The set of visual words weight          \\ \hline
        ~$Sim_{\mathcal{I}}(\mathcal{I}_{i}(\mathcal{W}_i),\mathcal{I}_{j}(\mathcal{W}_j))$    &The image similarity measurement        \\ \hline
        ~$\mu_i$                                         & The similarity of visual word                  \\ \hline
	\end{tabular}
    \caption{The summary of notations} \label{tab:notation}	
\end{table}

\subsection{Problem Definition}
\begin{definition}[\textbf{Image object}] \label{def:imgobj}
Let $\mathcal{D_\mathcal{I}}$ be an image dataset and $\mathcal{I}_i$ and $\mathcal{I}_j$ be two images, $\mathcal{I}_i, \mathcal{I}_j \in \mathcal{D}_\mathcal{I}$. We define the image object represented by bag-of-visual-word model as $\mathcal{I}_{i}(\mathcal{W}_i)$ and $\mathcal{I}_{j}(\mathcal{W}_j)$, wherein $\mathcal{W}_i=\{w^i_1,w^i_2,...,w^i_m\}$ and $\mathcal{W}_j= \{w^j_1,w^j_2,...,w^j_n\}$ are the visual word set generated by low-level feature extraction from $\mathcal{I}_i$ and $\mathcal{I}_j$, $|\mathcal{W}_i| = m$ and $|\mathcal{W}_j| = n$ are the number of visual words in these two sets respectively. In this study, we utilize image object as the representation model of images for the task of image similarity measurement.
\end{definition}

\begin{definition}[\textbf{Similarity of visual word}] \label{def:sim_vw}
Given two image objects $\mathcal{I}_{i}(\mathcal{W}_i)$ and $\mathcal{I}_{j}(\mathcal{W}_j)$, wherein $\mathcal{W}_i = \{w^i_1,w^i_2,...,w^i_m\}$ and $\mathcal{W}_j = \{w^j_1,w^j_2,...,w^j_n\}$ are the visual words set. The similarity of two visual word $w^i_k \in \mathcal{W}_i$ and $w^j_k \in \mathcal{W}_j$ is represented by $\lambda_k = Sim_\mathcal{W}(w^i_k, w^j_k), \lambda_k \in [0,1]$,  and if these visual words are identical, i.e., $\mathcal{W}_i = \mathcal{W}_j$, $\lambda_k = 1$.
\end{definition}

\begin{definition}[\textbf{Similar visual word pair}] \label{def:svwp}
Given two visual words $w^i_k \in \mathcal{W}_i$ and $w^j_k \in \mathcal{W}_j$ and the similarity of them is $\lambda_k = Sim_\mathcal{W}(w^i_k, w^j_k)$. Let $\bar{\lambda}$ is the similarity threshold predefined, if $\lambda_k > \bar{\lambda}$, this visual word pair is called as similar visual word pair (SVWP for short), represented as $\mathcal{P}_k=(w^i_k,w^j_k)$.
\end{definition}

\begin{definition}[\textbf{Similarity measurement of two image objects}] \label{def:sim_2img}
Given two image objects $\mathcal{I}_{i}(\mathcal{W}_i)$ and $\mathcal{I}_{j}(\mathcal{W}_j)$. Let operation $\mathcal{W}_i \bigotimes \mathcal{W}_j = \{\mathcal{P}_1,\mathcal{P}_2,...,\mathcal{P}_l\}$ generates the set of SVWPs which contain the visual words in $\mathcal{W}_i$ and $\mathcal{W}_j$, $l=|\mathcal{W}_i \bigotimes \mathcal{W}_j|$, and the similarity set of them are denoted as $\Lambda = \{\lambda_1,\lambda_2,...,\lambda_l\}$, $\forall \lambda_i \in \Lambda, \lambda_i > \bar{\lambda}$. Let $\xi^i_k$ and $\xi^j_k$ be the weight of visual word $w^i_k$ and $w^j_k$. For image objects $\mathcal{I}_{i}(\mathcal{W}_i)$ and $\mathcal{I}_{j}(\mathcal{W}_j)$, the sets of their visual words weight are denoted as $\Xi_i = \{\xi^i_1,\xi^i_2,...,\xi^i_l\}$ and $\Xi_j = \{\xi^j_1,\xi^j_2,...,\xi^j_l\}$. The definitional equation of similarity between $\mathcal{I}_{i}(\mathcal{W}_i)$ and $\mathcal{I}_{j}(\mathcal{W}_j)$ is shown as follows:
\begin{equation}\label{equ:sim_2img}
Sim_{\mathcal{I}}(\mathcal{I}_{i}(\mathcal{W}_i),\mathcal{I}_{j}(\mathcal{W}_j)) = \mathcal{F}(m,n,l,\Lambda,\Xi_i,\Xi_j)
\end{equation}
where $m$ and $n$ are the number of visual words of $\mathcal{I}_{i}(\mathcal{W}_i)$ and $\mathcal{I}_{j}(\mathcal{W}_j)$ respectively. It is clearly that $Sim_{\mathcal{I}}(\mathcal{I}_{i}(\mathcal{W}_i),\mathcal{I}_{j}(\mathcal{W}_j))$ can meet the systematic similarity measurement criterion.
\end{definition}

\begin{theorem}[\textbf{Monotonicity of similarity function}] \label{thm:monoton}
The similarity measurement $Sim_{\mathcal{I}}(\mathcal{I}_{i}(\mathcal{W}_i),\mathcal{I}_{j}(\mathcal{W}_j))$ has the following five monotonicity conditions:
\begin{itemize}
  \item  $Sim_{\mathcal{I}}(\mathcal{I}_{i}(\mathcal{W}_i),\mathcal{I}_{j}(\mathcal{W}_j))$ is a monotonic increasing function of weights of visual words in SVWPs, i.e., $\forall \xi_{w^i_x} \in \Xi_i$ and $\xi_{w^j_y} \in \Xi_j$, and $\forall \xi_{\hat{w}^i_x} \in \hat{\Xi}_i$ and $\xi_{\hat{w}^j_y} \in \hat{\Xi}_j$, if $\xi_{w^i_x} > \xi_{\hat{w}^i_x}$ and $\xi_{w^j_y} > \xi_{\hat{w}^j_y}$, $\mathcal{F}(m,n,l_1,\Lambda,\Xi_i,\Xi_j)>\mathcal{F}(m,n,l_2,\Lambda,\hat{\Xi}_i,\hat{\Xi}_j)$.
  \item  $Sim_{\mathcal{I}}(\mathcal{I}_{i}(\mathcal{W}_i),\mathcal{I}_{j}(\mathcal{W}_j))$ is a monotonic increasing function of the similarities of SVMPs $\Lambda = \{\lambda_1,\lambda_2,...\lambda_l\}$, i.e., $\forall \lambda_x \in \Lambda$ and $\hat{\lambda_x} \in \hat{\Lambda}$, $\mathcal{F}(m,n,l_1,\Lambda,\Xi_i,\Xi_j)>\mathcal{F}(m,n,l_2,\hat{\Lambda},\Xi_i,\Xi_j)$.
  \item  $Sim_{\mathcal{I}}(\mathcal{I}_{i}(\mathcal{W}_i),\mathcal{I}_{j}(\mathcal{W}_j))$ is a monotonic increasing function of number of SVWPs $l$, i.e., $\forall l_1,l_2 \in \textbf{N}^+$, if $l_1>l_2$, $\mathcal{F}(m,n,l_1,\Lambda,\Xi_i,\Xi_j)>\mathcal{F}(m,n,l_2,\Lambda,\Xi_i,\Xi_j)$.
  \item  $Sim_{\mathcal{I}}(\mathcal{I}_{i}(\mathcal{W}_i),\mathcal{I}_{j}(\mathcal{W}_j))$ is a monotonic decreasing function of weights of visual words which are not in SVWPs, i.e., .
  \item  $Sim_{\mathcal{I}}(\mathcal{I}_{i}(\mathcal{W}_i),\mathcal{I}_{j}(\mathcal{W}_j))$ is a monotonic decreasing function of the number of visual words which are not in SVWPs, i.e., if $r_1=m+n-l_1$ and $r_2=m+n-l_2$, $r_1 > r_2 \to l_1 < l_2$, $\mathcal{F}(m,n,l_1,\Lambda,\Xi_i,\Xi_j)<\mathcal{F}(m,n,l_2,\Lambda,\Xi_i,\Xi_j)$.
\end{itemize}
\end{theorem}

According to the Definition~\ref{def:sim_2img} and theorem~\ref{thm:monoton}, the similarity measurement for two image objects is proposed, which is described in formal as follows.

Given two image objects $\mathcal{I}_{i}(\mathcal{W}_i)$ and $\mathcal{I}_{j}(\mathcal{W}_j)$, $m=|\mathcal{W}_i|$ and $n=|\mathcal{W}_j|$. The sets of their visual words weight are $\Xi_i = \{\xi^i_1,\xi^i_2,...\xi^i_l\}$ and $\Xi_j = \{\xi^j_1,\xi^j_2,...\xi^j_l\}$. The SVMPs set of $\mathcal{I}_{i}(\mathcal{W}_i)$ and $\mathcal{I}_{j}(\mathcal{W}_j)$ is $\{\mathcal{P}_1,\mathcal{P}_2,...,\mathcal{P}_l\}$, $l \leq min(m,n)$, and the similarities set of them is $\Lambda = \{\lambda_1,\lambda_2,...\lambda_l\}$. The similarity measurement function $Sim_{\mathcal{I}}(\mathcal{I}_{i}(\mathcal{W}_i),\mathcal{I}_{j}(\mathcal{W}_j))$ is:

\begin{equation}\label{equ:sim_func}
Sim_{\mathcal{I}}(\mathcal{I}_{i}(\mathcal{W}_i),\mathcal{I}_{j}(\mathcal{W}_j)) = \frac{\sum\limits_{k=1}^{l}\lambda_k\xi^i_k\xi^j_k}{\sqrt{\sum\limits_{k=1}^{m}\xi^i_k\sum\limits_{k=1}^{n}\xi^j_k}
\sqrt{\sum\limits_{k=1}^{l}{\lambda_k}^2\xi^i_k\xi^j_k+\sum\limits_{k=l+1}^{m}\xi^i_k\sum\limits_{k=l+1}^{n}\xi^j_k}}
\end{equation}

Function~\ref{equ:sim_func} apparently meet the monotonicity described in Theorem~\ref{thm:monoton}. On the other hand, if these two image objects are identical, i.e., $\mathcal{I}_{i}(\mathcal{W}_i) = \mathcal{I}_{j}(\mathcal{W}_j)$, $\mathcal{W}_i = \mathcal{W}_j$, $m=n=l$, and $\xi^i_k=\xi^j_k$, then $Sim_{\mathcal{I}}(\mathcal{I}_{i}(\mathcal{W}_i),\mathcal{I}_{j}(\mathcal{W}_j)) = 1$.

\begin{theorem}[\textbf{dissatisfying commutative law}] \label{thm:commutative}
The similarity measurement $Sim_{\mathcal{I}}(\mathcal{I}_{i}\\(\mathcal{W}_i),\mathcal{I}_{j}(\mathcal{W}_j))$ dissatisfy commutative law, i.e.,
\begin{equation*}
Sim_{\mathcal{I}}(\mathcal{I}_{i}(\mathcal{W}_i),\mathcal{I}_{j}(\mathcal{W}_j)) \neq Sim_{\mathcal{I}}(\mathcal{I}_{j}(\mathcal{W}_j),\mathcal{I}_{i}(\mathcal{W}_i))
\end{equation*}
\end{theorem}

In general, some visual words (e.g., noise words) in image objects have negative or reverse effects on the expression of the whole image. The SMI has a penalty effect on non-similar visual elements according to Theorem~\ref{thm:monoton}. this feature of the SMI has high accuracy for the similarity measurement of images.

\section{Image Similarity Measurement Algorithm}
\label{optimization}

\subsection{The Measurement of Similar Visual Words}
SMI is subject to the time complexity of the calculation of similar visual words. $\mu_i$ represents the similarity of a similar visual word as shown in the following formula:
\begin{equation}\label{equ:calcvw}
\mu_i =
\begin{cases}
\mathop{\arg\max}_{b_j \in S_B}Sim_{\mathcal{I}}(a_i,b_j), \quad if > \mu_0 \\
0, \qquad  \qquad  \qquad  \qquad  \qquad \quad \quad  if < \mu_0
\end{cases}
\end{equation}
where $Sim_{\mathcal{I}}(a_i,b_j)$ represents the cosine of the angle between two vectors as the measurement of similarity. $\mu_0$ is a judgment of the similarity threshold.

We give an intuitive way to measure similar visual words. The pseudo-code of the algorithm is shown in Algorithm~\ref{alg:simmin}. In this work, the double loop cosine calculation method is called to be SMI Naive (SMIN for short).

\begin{algorithm}
\begin{algorithmic}[1]
%\small
%\scriptsize
\footnotesize
\caption{\bf SMIN Algorithm}
\label{alg:simmin}

\INPUT $S_A$, $S_B$, $\mu_0$.
\OUTPUT $\mu$.

\STATE Initializing: $\mu \leftarrow \emptyset$;
\STATE Initializing: $S \leftarrow \emptyset$;
\STATE Initializing: $N_S \leftarrow \emptyset$;
\STATE Initializing: $maxsim \leftarrow 0$;

\FOR{each $\mathcal{W}_i \in S_A$}
    \FOR{each $\mathcal{W}'_j \in S_B$}
        \IF{$cos(\mathcal{W}_i,\mathcal{W}'_j)$}
            \STATE $maxsim \leftarrow cos(\mathcal{W}_i,\mathcal{W}'_j)$;
        \ENDIF
        \IF{$maxsim \geq \mu_0$}
            \STATE $S.Add(\mathcal{W}_i)$;
            \STATE $\mu.Add(maxsim)$;
        \ELSE
            \STATE $NS.Add(\mathcal{W}_i)$;
            \STATE $\mu.Add(0)$;
        \ENDIF
    \ENDFOR
\ENDFOR
\RETURN $\mu$;
\end{algorithmic}
\end{algorithm}

\subsection{The Optimization of Calculating Similar Visual Words}
In the context of massive multimedia data, the multimedia retrieval system or image similarity measurement system requires an efficient sentence similarity measurement algorithm, the time complexity of the SMI focuses on the optimization of calculating similar visual words.

\noindent\textbf{SMI Temp Index.} To reduce the double loop cos calculation to 1 cycle, a further approach is to construct an index $\gamma_i$ of $S_B$ for each vector $a_i$ in $S_A$. According to experience, the dimension of the visual word vector is generally 200-300 dimensions to get better results.

\begin{figure*}
\newskip\subfigtoppskip \subfigtopskip = -0.1cm
%\newskip\subfigcapskip \subfigcapskip = -0.2cm
\begin{minipage}[b]{0.99\linewidth}
\begin{center}
   %  \centering
     \includegraphics[width=1\linewidth]{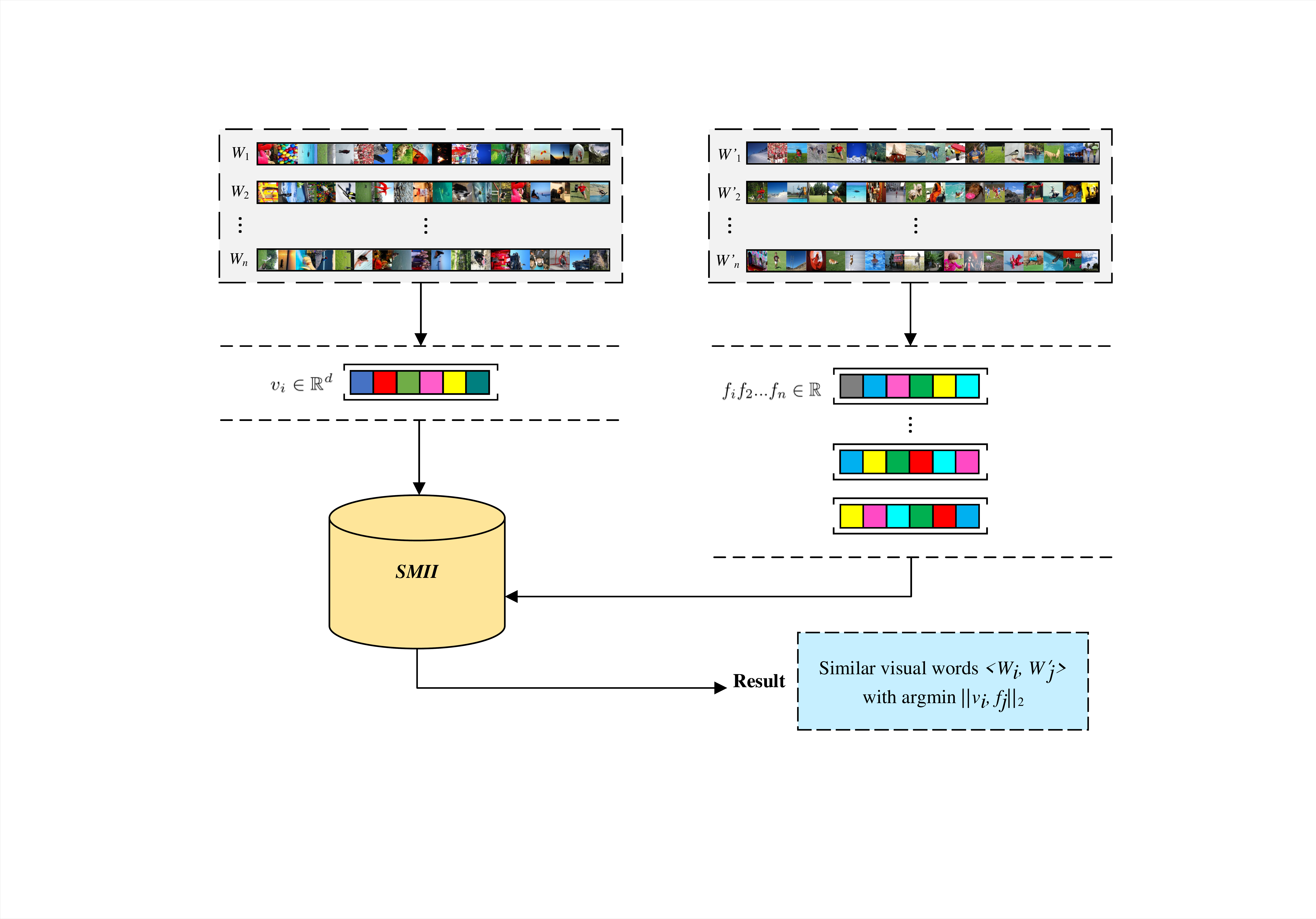}
   %\captionsetup{justification=centering,margin=2cm}
   \captionsetup{justification=centering}
       \vspace{-0.2cm}
\caption{The processing of similarity measurement via SMII}
\label{fig:fig3}
\end{center}
\end{minipage}
\label{fig:k}
\end{figure*}

For a vector $a_i$ in $S_A$, we search for the vector $b_j$ with the highest similarity in the temp index $\gamma_i$, so that the process requires only one similarity calculation. The $n$ times calculations of similar visual words $<a_i, b_j>$ are reduced to vector searching, thereby reducing the execution time of \emph{SMI}. However, there is a flaw that when every time a similar element of a sentence is calculated, a temp index needs to be built once, and the index cannot be reused. The temp index approach is called to be SMI Temp Index (SMII for short), as shown in Figure~\ref{fig:fig3}.

\noindent\textbf{Index of Potential Similar Visual Words.} In order to solve the problem that the index cannot be reused, we establish an index of potential similar visual words off-line in the process of word vector training. We could search for the index to perform the measurement of similar visual words without having to repeatedly create a temporary index. The main steps for index of potential similar visual words construction is shown as follows:
\begin{itemize}
\item Establishing an index for all the visual word vector set by trained word vector model.
\item Traversing any vector $\mathfrak{v}$ to search the index to get a return set. In this set, the potential similar visual words are abstained with the similarity is greater than the threshold $\mu_0$, in similarity descending order.
\item The physic indexing structure of potential similar visual words could be implemented by a Huffman tree.
\end{itemize}

According to the hierarchical Softmax strategy in Word2Vec, an original Word2Vec Huffman tree constructed on the basis of the visual words frequency, and each node (except the root node) represents a visual word and its corresponding vector.

We try to replace the vector with potential similar visual words. Thus each node of tree represents a visual word and its corresponding potential similar visual words. The index structure is illustrated by Figure~\ref{fig:fig4}:

\begin{figure*}
\newskip\subfigtoppskip \subfigtopskip = -0.1cm
%\newskip\subfigcapskip \subfigcapskip = -0.2cm
\begin{minipage}[b]{0.99\linewidth}
\begin{center}
   %  \centering
     \includegraphics[width=1\linewidth]{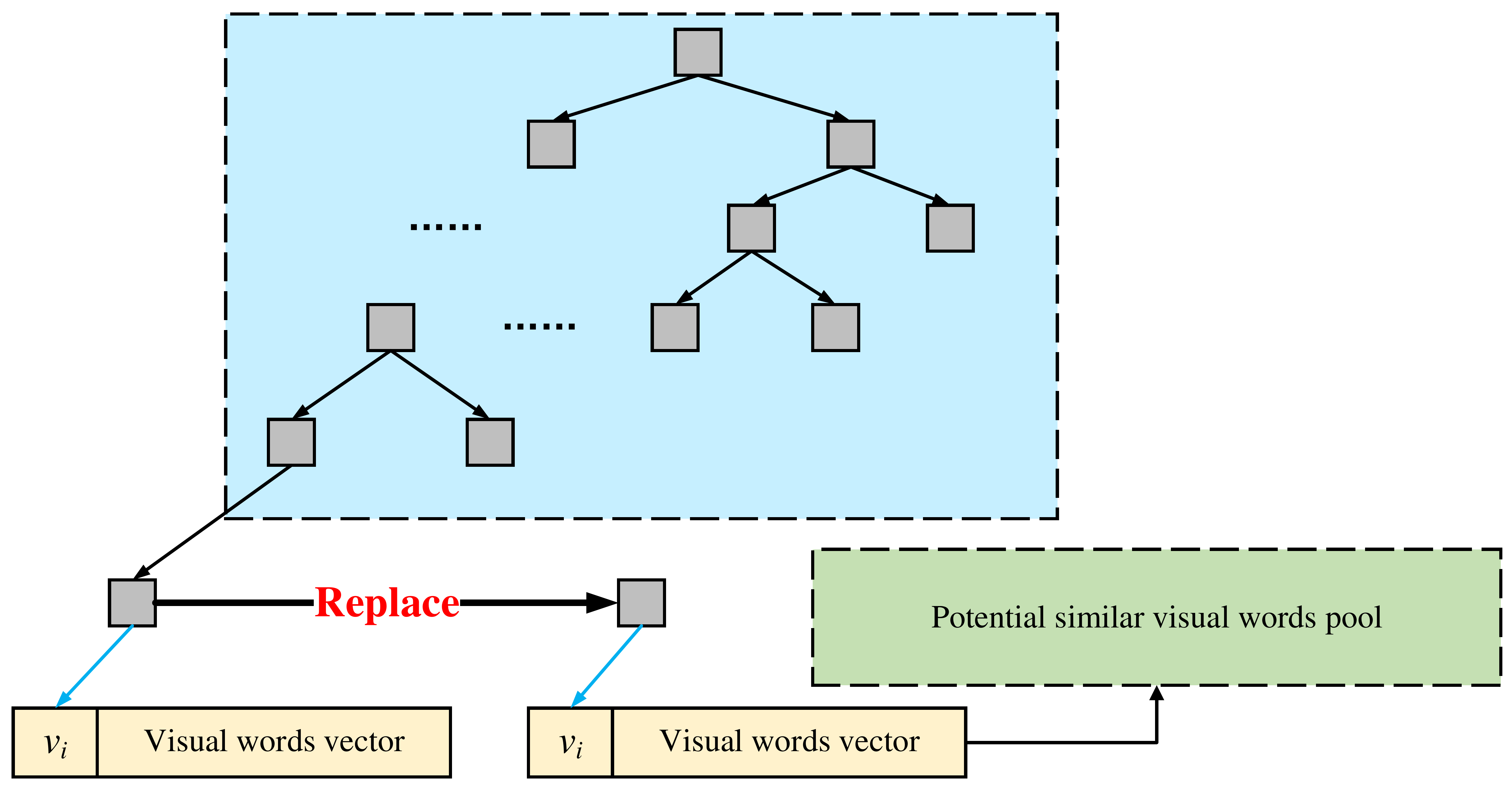}
   %\captionsetup{justification=centering,margin=2cm}
   \captionsetup{justification=centering}
       \vspace{-0.2cm}
\caption{The index structure of potential similar visual words}
\label{fig:fig4}
\end{center}
\end{minipage}
\label{fig:k}
\end{figure*}

We call the methods using global index of potential similar visual words as PSMI. Algorithm~\ref{alg:psmii} illustrates the pseudo-code of PSMI.

\begin{algorithm}
\begin{algorithmic}[1]
%\small
%\scriptsize
\footnotesize
\caption{\bf PSMI Algorithm}
\label{alg:psmii}

\INPUT  $S_A$, $S_B$, $\mu_0$
\OUTPUT $\mu$.

\STATE Initializing: $\mu \leftarrow \emptyset$;
\STATE Initializing: $S \leftarrow \emptyset$;
\STATE Initializing: $NS \leftarrow \emptyset$;
\STATE Initializing: $\mathcal{P} \leftarrow \emptyset$;
\STATE Initializing: $maxsim \leftarrow 0$;

\FOR{each $\mathcal{W}_i \in S_A$}
    \STATE $\mathcal{P} \leftarrow HuffmanSearch(\mathcal{W}_i)$;
    \FOR{each $\mathcal{W}'_j \in S_B$}
        \FOR{each $p_k \in \mathcal{P}$}
            \IF{$\mathcal{W}_j.equal(p_k.vector)$}
                \STATE $S.Add(\mathcal{W}_i)$;
                \STATE $\mu.Add(p_k.sim)$;
                \STATE Break to loop $\mathcal{W}_i$;
            \ENDIF
        \ENDFOR
    \ENDFOR
    \STATE $NS.Add(\mathcal{W}_i)$;
    \STATE $\mu.Add(0)$;
\ENDFOR
\RETURN $\mu$;
\end{algorithmic}
\end{algorithm}

Algorithm~\ref{alg:psmii} demonstrates the processing of the PSMI Algorithm. Firstly, for each visual words vector $\mathcal{W}_i \in S_A$, the algorithm executes the procedure $HuffmanSearch(\mathcal{W}_i)$ to get the node of the huffman tree which contains $\mathcal{W}_i$ and stored it in $\mathcal{P}$. Then, for each $\mathcal{W}'_j \in S_B$, the algorithm select each $p_k$ from $\mathcal{P}$ and check if $\mathcal{W}_j$ is equal to $p_k.vector$ or not. if them are equal, the algorithm adds $\mathcal{W}_i$ into set $S$ and adds $p_k.sim$ into $\mu$. Then break to the outer loop. If $\mathcal{W}_i$ and $p_k.vector$ are not equal, then adds $\mathcal{W}_i$ into set $NS$ and add 0 into $\mu$.

\section{PERFORMANCE EVALUATION}
\label{perform}

In this section, we present results of a comprehensive performance study on real
image datasets Flickr and ImageNet to evaluate the efficiency and scalability of the proposed techniques.
Specifically, we evaluate the effectiveness of the following indexing techniques for region of visual interests search on road network.
\begin{itemize}
\item{\textbf{WJ}} WJ is the word2Vec technique proposed in \url{https://github.com/jsksxs360/Word2Vec}.
\item{\textbf{WMD}} WMD is the word2Vec technique, which is based on moving distance, is proposed in \url{https://github.com/crtomirmajer/wmd4j}.
\item{\textbf{SMIN}} SMIN is the double loop cosine calculation technique proposed in Section ~\ref{optimization}.
\item{\textbf{SMII}} SMII is the advanced technique of SMIN, which is proposed in Section ~\ref{optimization}.
\item{\textbf{PSMI}} PSMI is the potential similar visual words technique of SMII, which is also proposed in Section ~\ref{optimization}.
\end{itemize}

\noindent \textbf{Datasets.} Performance of various algorithms is evaluated on both real image datasets.

We first evaluate these algorithms on \textbf{Flickr} is obtained by crawling millions image the photo-sharing site Flickr(\url{http://www.flickr.com/}). For the scalability and performance evaluation, we randomly sampled five sub datasets whose sizes vary from 200,000 to 1000,000 from the image dataset. Similarly, another image dataset \textbf{ImageNet}, which is widely used in image processing and computer vision, is used to evaluate the performance of these algorithms. Dataset \textbf{ImageNet} not only includes 14,197,122 images, but also contained 1.2 million images with SIFT features.  We generate \textbf{ImageNet} dataset with varying size from 20K to 1M.

{\bf Workload.}
A workload for the region of visual interests query consists of $100$ queries. The accuracy of these algorithm and the query response time is employed to evaluate the performance of the algorithms. The image dataset size grows from 0.2M to 1M; the number of the query visual words of dataset \textbf{Flickr} changes from 20 to 100; the number of the query visual words of dataset \textbf{ImageNet} varies from 50 to 250. The image dataset size, the number of the query visual words of dataset \textbf{Flickr}, and the number of the query visual words of dataset \textbf{ImageNet} set to 0.2M, 40, 100 respectively. Experiments are run on a PC with Intel Xeon 2.60GHz dual CPU and 16G memory running Ubuntu. All algorithms in the experiments are implemented in Java. Note that we only consider the algorithms WJ, SMI, WDM in accuracy comparison, because the SMIN, SMII, PSMI algorithms have the same error tolerance.

\begin{figure*}
\newskip\subfigtoppskip \subfigtopskip = -0.1cm
%\newskip\subfigcapskip \subfigcapskip = -0.2cm
\begin{minipage}[b]{1\linewidth}
\begin{center}
   %  \centering
     \subfigure[Evaluation on Flickr]{
     \includegraphics[width=0.48\linewidth]{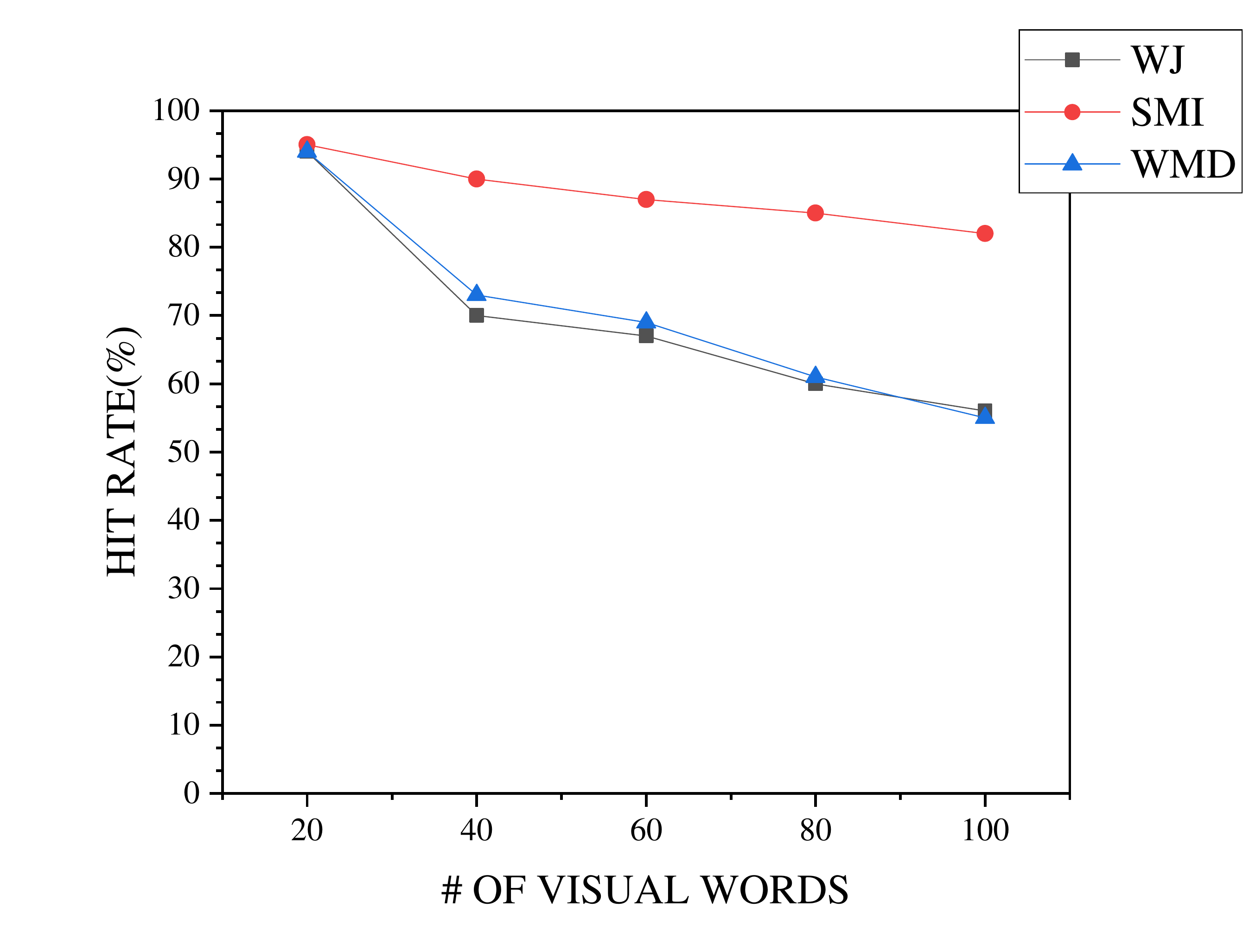}
     }
     \subfigure[Evaluation on ImageNet]{
     \includegraphics[width=0.48\linewidth]{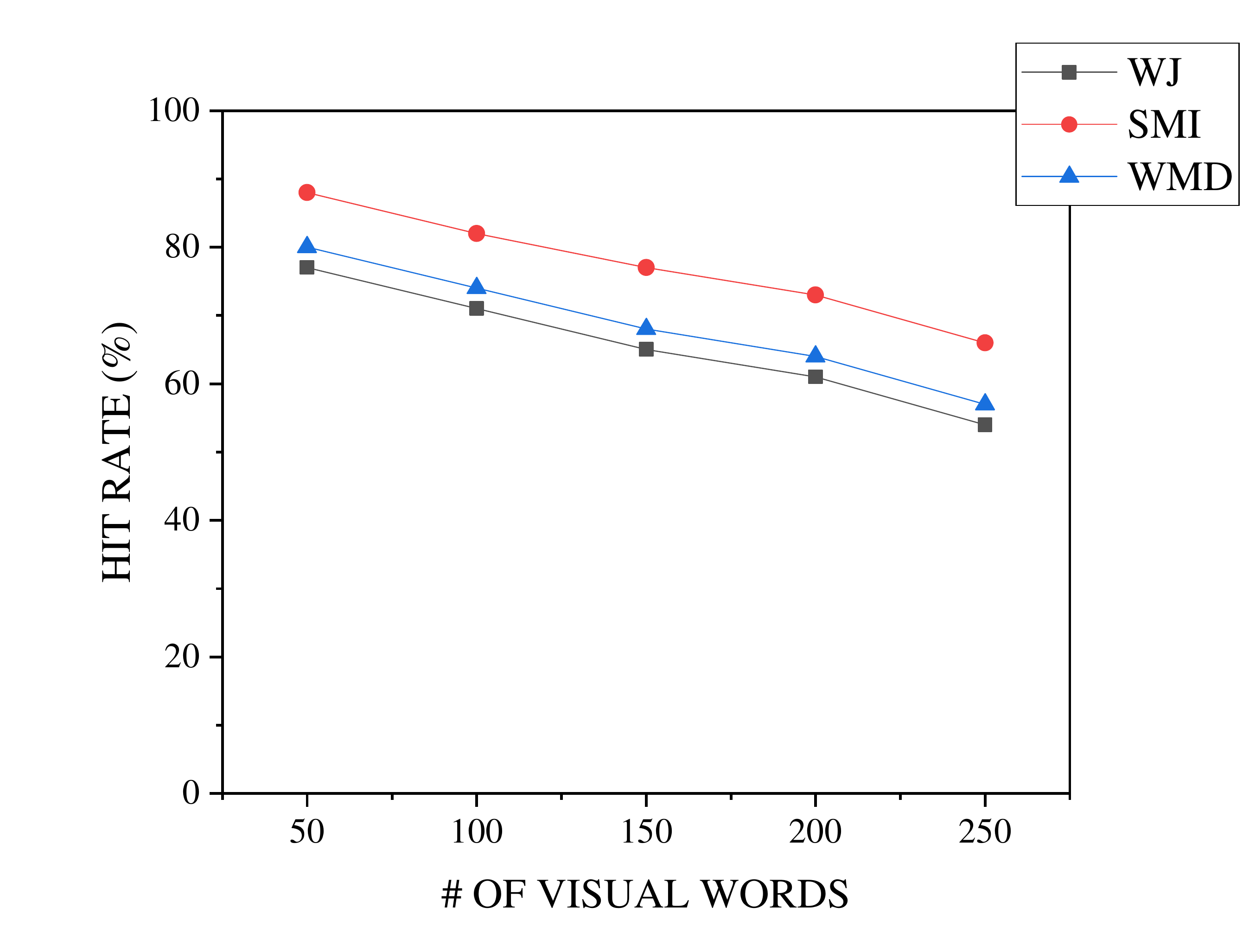}
     }
   %\captionsetup{justification=centering,margin=2cm}
   \captionsetup{justification=centering}
       \vspace{-0.2cm}
\caption{Evaluation on the number of visual words on Flickr and ImageNet}
\label{fig:number-of-visual-words}
\end{center}
\end{minipage}
\end{figure*}

\begin{figure*}
\newskip\subfigtoppskip \subfigtopskip = -0.1cm
%\newskip\subfigcapskip \subfigcapskip = -0.2cm
\begin{minipage}[b]{1\linewidth}
\begin{center}
   %  \centering
     \subfigure[Evaluation on Flickr]{
     \includegraphics[width=0.48\linewidth]{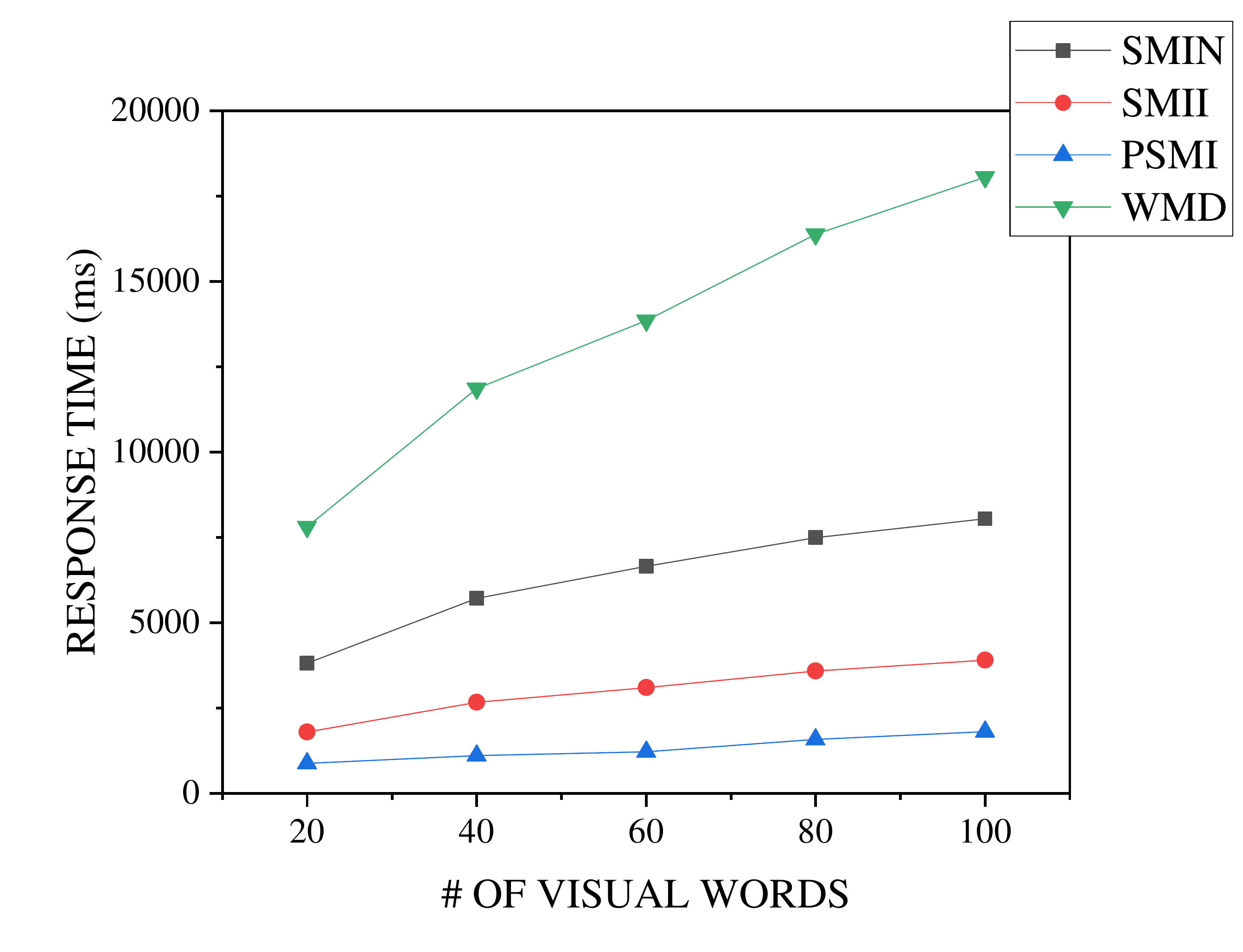}
     }
     \subfigure[Evaluation on ImageNet]{
     \includegraphics[width=0.48\linewidth]{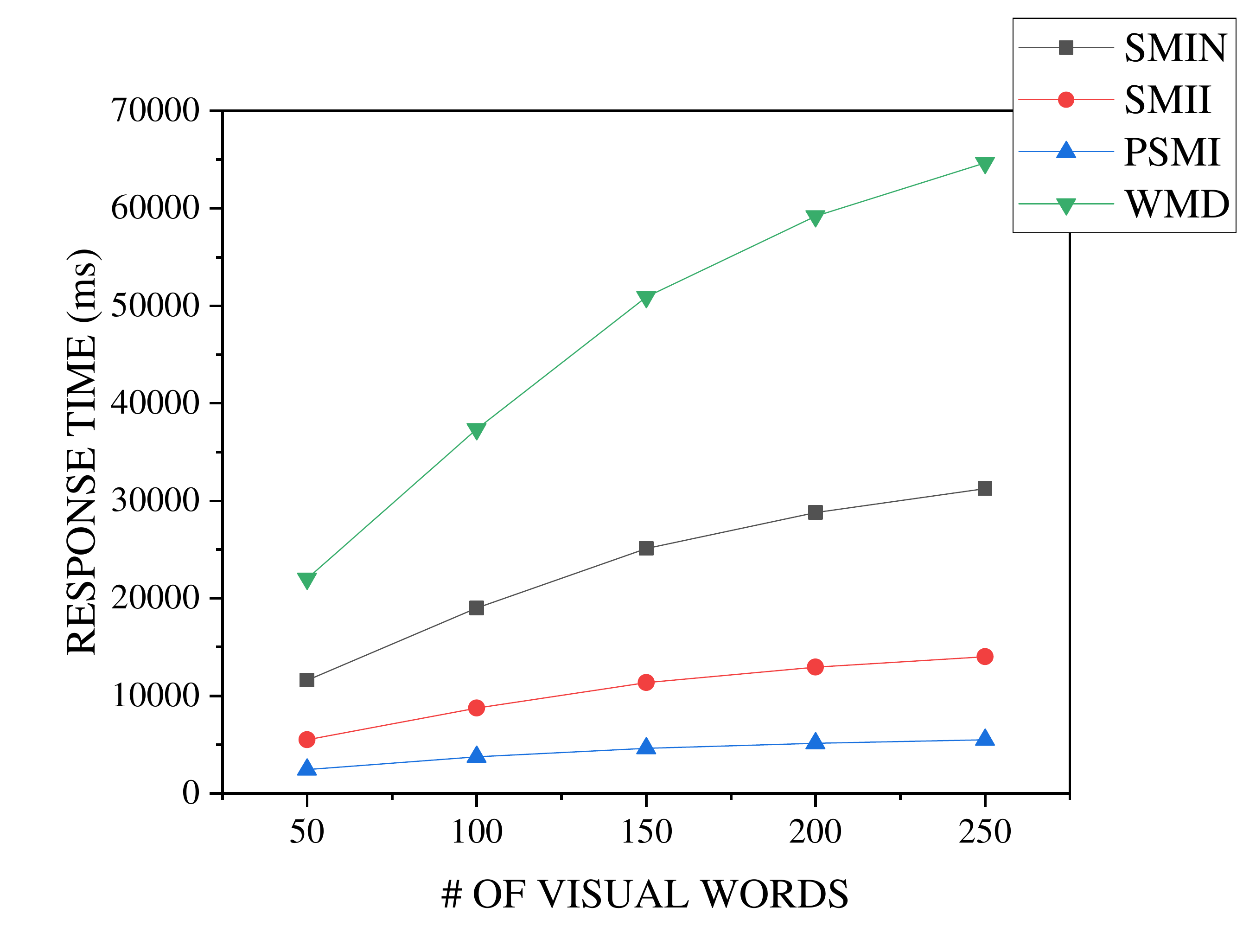}
     }
   %\captionsetup{justification=centering,margin=2cm}
   \captionsetup{justification=centering}
       \vspace{-0.2cm}
\caption{Evaluation on the number of visual words on Flickr and ImageNet}
\label{fig:number-of-visual-words-response}
\end{center}
\end{minipage}
\end{figure*}

\noindent \textbf{Evaluating hit rate on the number of visual words.}  We evaluate the hit rate on the number of query visual words on Flickr and ImageNet dataset shown in Figure~\ref{fig:number-of-visual-words}. The experiment on Flickr is shown in Figure~\ref{fig:number-of-visual-words}(a). It is clear that the hit rate of WJ, SMI and WMD decrease with the rising of the number of visual words. Specifically, the hit rate of our method, SMI, is the highest all the time. It descends slowly from around 90\% to about 85\%. On the other hand, the hit rate of WJ and WMD are very close. In the interval $[20,40]$, they go down rapidly and after that the decrement of them become moderate. At 100, the hit rate of WJ is a litter higher than WMD, and both of them are much lower than SMI. In Figure~\ref{fig:number-of-visual-words}(b), all of the decreasing trends are similar. Apparently, the hit rate of SMI is the highest, which goes down gradually with the increasing of the number of visual words. On ImageNet dataset, the hit rate of WMD is a litter higher than WJ all the time.

\noindent \textbf{Evaluating response time on the number of visual words.}  We evaluate the response time on the number of visual words on Flickr and ImageNet dataset shown in Figure~\ref{fig:number-of-visual-words-response}. In Figure~\ref{fig:number-of-visual-words-response}(a), with the increment of number of visual words, the response time of PSMI has a slight growth, which is the lowest in these methods. The increasing trends of SMII is very moderate too, but it is slightly inferior to PSMI. Like PSMI and SMII, the performance of SMIN shows a moderate decrement with the rising of spatial similarity threshold. Although the response time of it is higher than the former two, it is much lower than WMD which has a fast growth in the interval of $20,100$. Figure~\ref{fig:number-of-visual-words-response}(b) illustrates that the efficiency of PSMI is almost the same with the increment of number of visual words, which is the highest amount these four methods. Like the experiment on Flickr, the performance of both SMII and SMIN increase gradually and they are much better than WMD.

\begin{figure*}
\newskip\subfigtoppskip \subfigtopskip = -0.1cm
%\newskip\subfigcapskip \subfigcapskip = -0.2cm
\begin{minipage}[b]{1\linewidth}
\begin{center}
   %  \centering
     \subfigure[Evaluation on Flickr]{
     \includegraphics[width=0.48\linewidth]{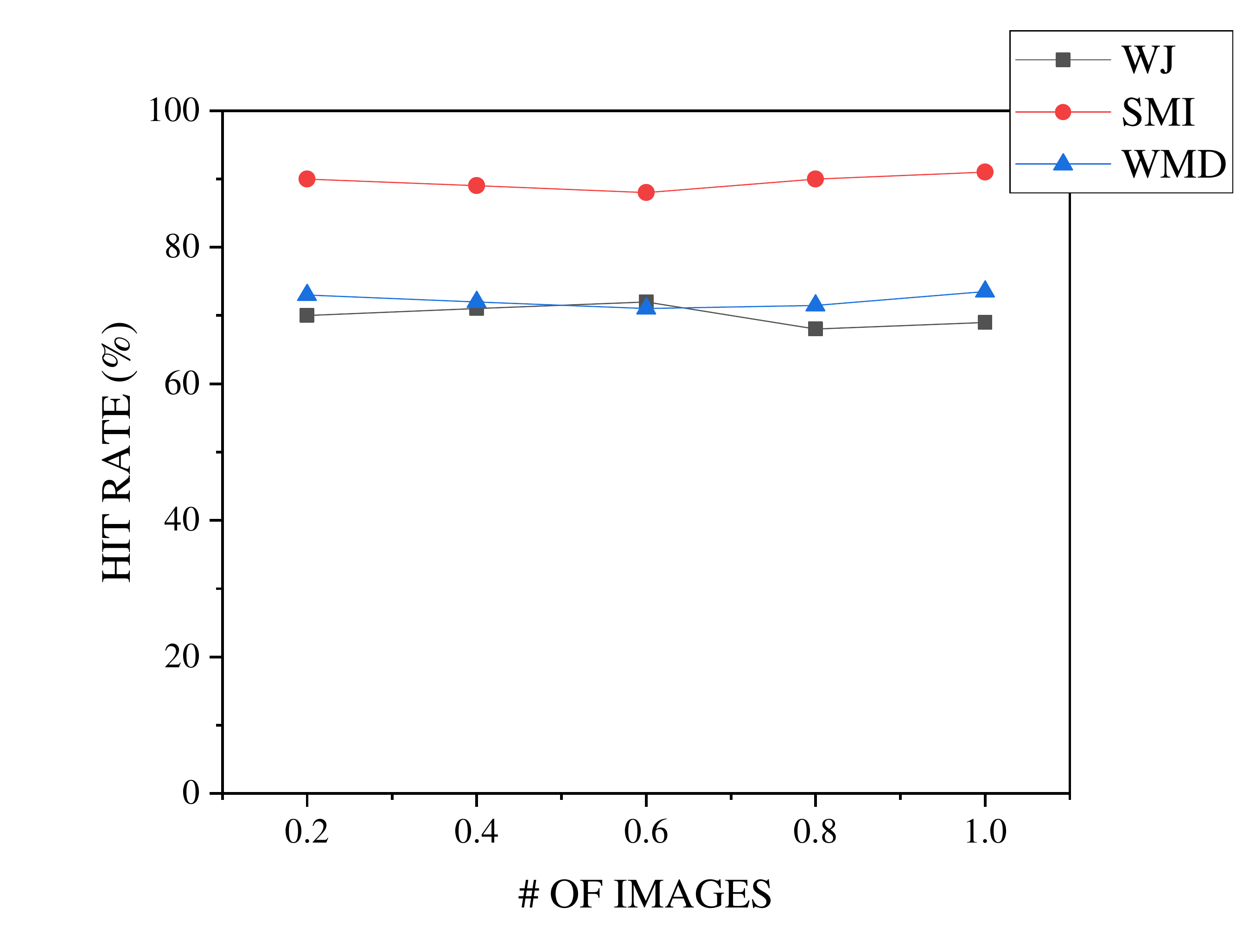}
     }
     \subfigure[Evaluation on ImageNet]{
     \includegraphics[width=0.48\linewidth]{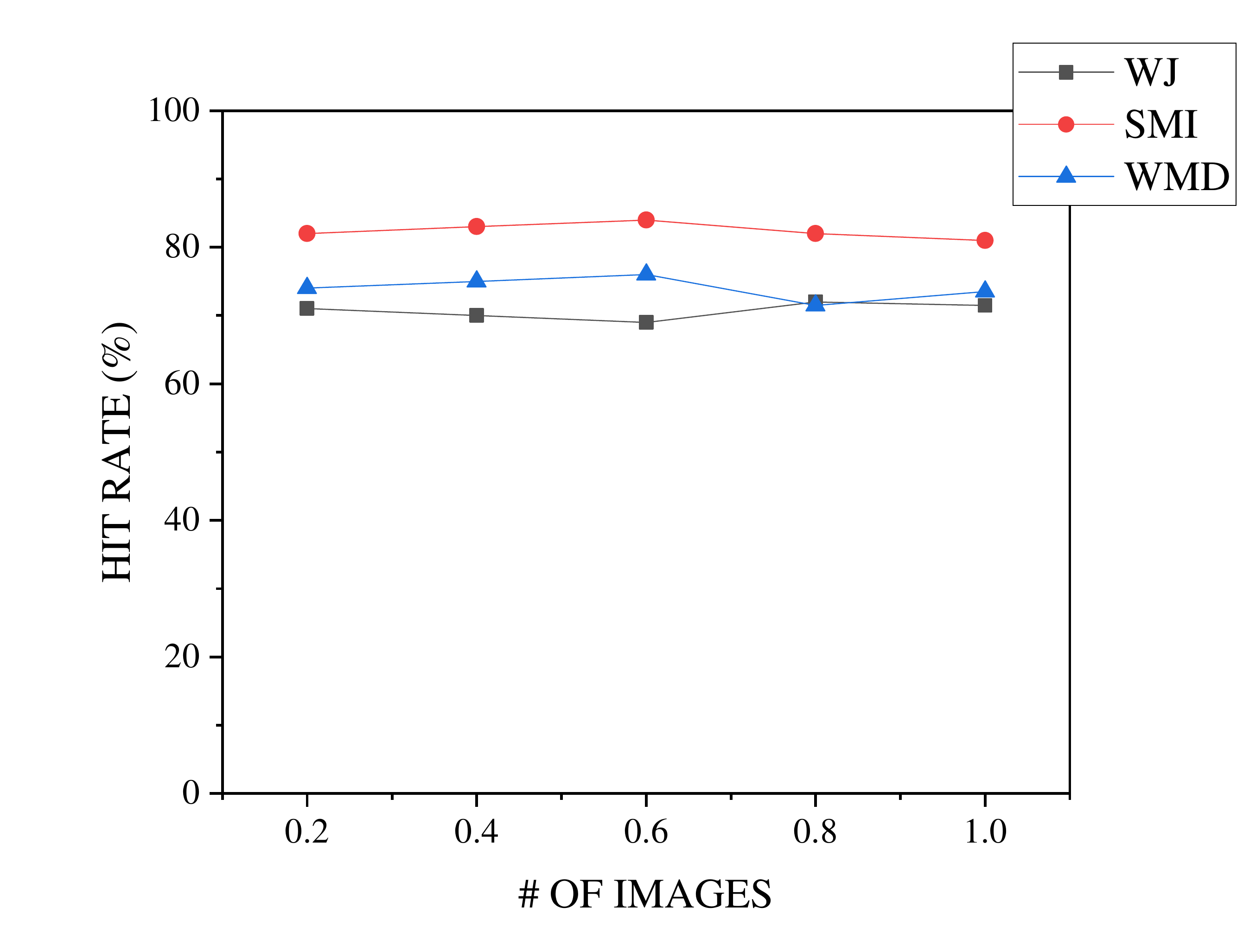}
     }
   %\captionsetup{justification=centering,margin=2cm}
   \captionsetup{justification=centering}
       \vspace{-0.2cm}
\caption{Evaluation on the number of images on Flickr and ImageNet}
\label{fig:number-of-visual-words}
\end{center}
\end{minipage}
\end{figure*}

\begin{figure*}
\newskip\subfigtoppskip \subfigtopskip = -0.1cm
%\newskip\subfigcapskip \subfigcapskip = -0.2cm
\begin{minipage}[b]{1\linewidth}
\begin{center}
   %  \centering
     \subfigure[Evaluation on Flickr]{
     \includegraphics[width=0.48\linewidth]{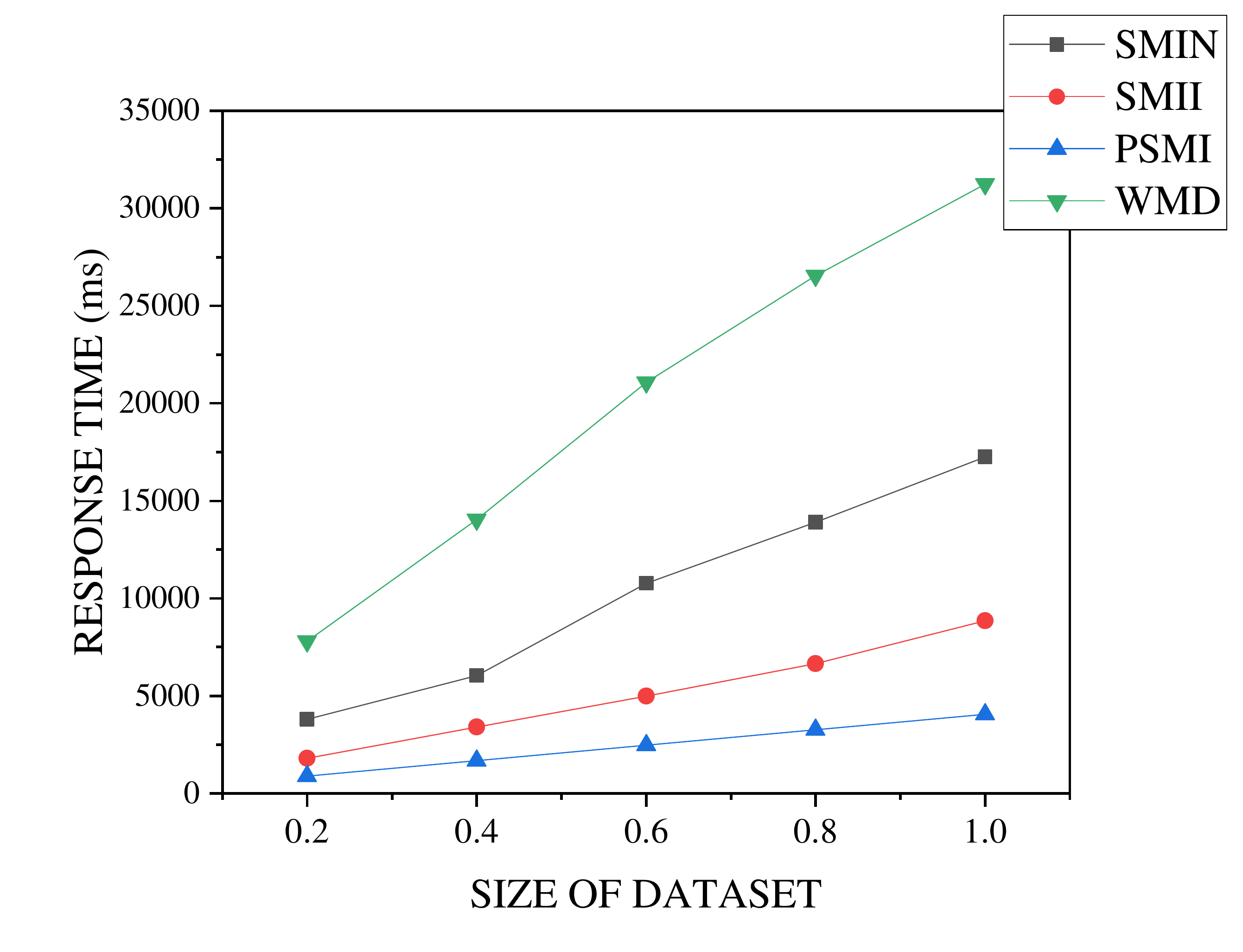}
     }
     \subfigure[Evaluation on ImageNet]{
     \includegraphics[width=0.48\linewidth]{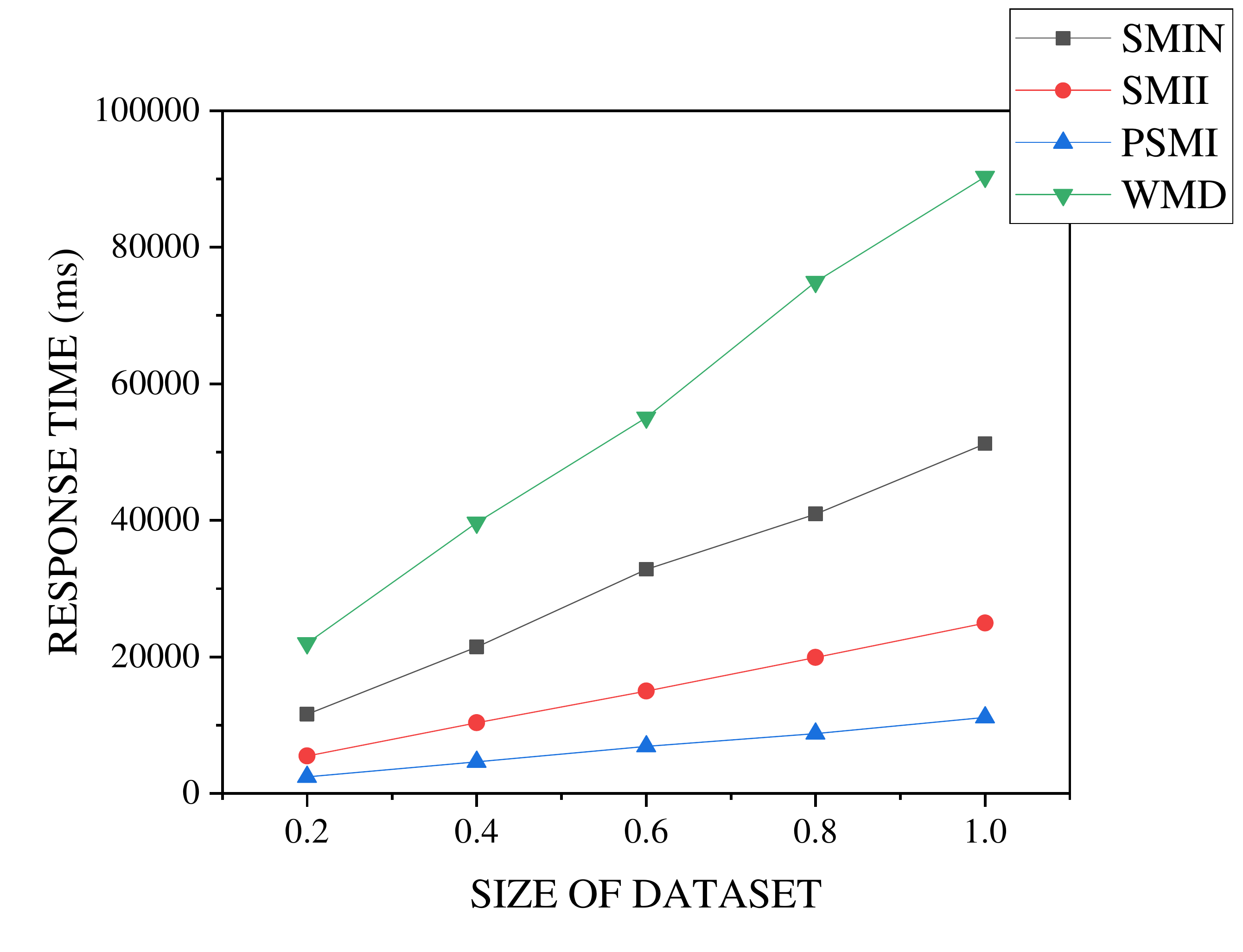}
     }
   %\captionsetup{justification=centering,margin=2cm}
   \captionsetup{justification=centering}
       \vspace{-0.2cm}
\caption{Evaluation on the number of images on Flickr and ImageNet}
\label{fig:number-of-visual-words-response}
\end{center}
\end{minipage}
\end{figure*}

\noindent \textbf{Evaluating hit rate on the number of images.}  We evaluate the hit rate on the number of images on Flickr and ImageNet dataset shown in Figure~\ref{fig:number-of-visual-words}. Figure~\ref{fig:number-of-visual-words}(a) demonstrates clearly that the hit rate of SMI is much higher than WJ and WMD. With the increasing of images number, it fluctuates slightly. the hit rate of WMD is almost unchanged with the increasing of number of images. On the other hand, the hit rate of WJ shows a moderate growth in the interval of $0.2,0.6$ and after that it drops and it is a litter lower than WMD. Clearly, the performance of SMI is the best. Figure~\ref{fig:number-of-visual-words}(b) shows that the hit rate of SMI grows slightly in $[0.2,.06]$ and then go down weakly, which is higher than two others. Like the trend of SMI, the hit rate of WMD hit the maximum value at 0.6 and after that it decreases in the interval of $[0.6,0.8]$. With this just the opposite is that the hit rate of WJ has a moderate decrement in $[0.2,0.6]$ and it rises after 0.6.

\noindent \textbf{Evaluating response time on the number of images.} We evaluate response time on different size of query region on Flickr and ImageNet dataset shown in Figure~\ref{fig:number-of-visual-words-response}. We can find from Figure~\ref{fig:number-of-visual-words-response}(a) that the response time of PSMI and SMII increase slowly with the increasing of size of dataset. Both of them are much better than the others. The growth rate of SMIN is a litter higher than the two formers. The efficiency time of WMD is the worst. It grows rapidly and at 1.0 it is more than 30000ms. In Figure~\ref{fig:number-of-visual-words-response}(b), we see that the growth of WMD is the fastest too. Like the situation on Flickr, the performance of WMD is the worst among them. By comparison, the upward trends of SMII and PSMI are much more moderate, and PSMI shows the best performance. 
\section{Conclusion}
\label{con}

In this paper, we investigate the problem of image similarity measurement which is a significant issue in many applications. Firstly we proposed the definition of image objects and similarity measurement of two images and related notions. We present the basic method of image similarity measurement which is named SMIN based on Word2Vec. To improve the performance of similarity calculation, we improve this method and propose SMI Temp Index. To solve the problem of that the index cannot be reused, we develop a novel indexing technique called Index of Potential Similar Visual Words (PSMI). The experimental evaluation on real geo-multimedia dataset shows that our solution outperforms the state-of-the-art method.

\textbf{Acknowledgments:} This work was supported in part by the National Natural Science Foundation of China
(61702560), project (2018JJ3691, 2016JC2011) of Science and Technology Plan of Hunan Province, and the Research and Innovation Project of Central South University Graduate Students(2018zzts177,2018zzts588).

\bibliographystyle{spmpsci}      % mathematics and physical sciences
\bibliography{ref}

\end{document}